\documentclass[aps,prd,prev,twocolumn,superscriptaddress,floatfix,nofootinbib]{revtex4-1}
\pdfoutput=1
\usepackage{graphicx}
\usepackage{bm}
\usepackage{times}
\usepackage{slashed}
\usepackage{color}
\usepackage{aas_macros}
\usepackage{slashed}
\usepackage{lipsum}
\usepackage{subfigure}
\usepackage{multirow}
\usepackage{amsmath}
\usepackage{array}
\usepackage{varwidth} 
\usepackage{hyperref}
\hypersetup{colorlinks = true, 
citecolor=red,
linkcolor=red,
filecolor=blue,      
urlcolor=blue
}

\newcommand{\ko}{kr_{L,0}}

\begin{document}
\title{Self-generated cosmic-Ray confinement in TeV halos:\\ Implications for TeV $\gamma$-ray emission and the positron excess}

\author{Carmelo Evoli} 
\email{carmelo.evoli@gssi.it}
\affiliation{Gran Sasso Science Institute (GSSI), Viale Francesco Crispi 7, 67100 L'Aquila, Italy}
\affiliation{INFN, Laboratori Nazionali del Gran Sasso (LNGS), 67100 Assergi, L'Aquila, Italy}

\author{Tim Linden}
\email{linden.70@osu.edu}
\affiliation{Center for Cosmology and AstroParticle Physics (CCAPP), and \\ Department of Physics, The Ohio State University Columbus, Ohio, 43210, USA}

\author{Giovanni Morlino} 
\email{giovanni.morlino@gssi.it}
\affiliation{Gran Sasso Science Institute (GSSI), Viale Francesco Crispi 7, 67100 L'Aquila, Italy}
\affiliation{INFN, Laboratori Nazionali del Gran Sasso (LNGS), 67100 Assergi, L'Aquila, Italy}
\affiliation{INAF/Osservatorio Astrofico di Arcetri, Largo E. Fermi 5, Firenze, Italy}

\begin{abstract}
\noindent Recent observations have detected extended TeV $\gamma$-ray emission surrounding young and middle-aged pulsars. The morphology of these ``TeV halos" requires cosmic-ray diffusion to be locally suppressed by a factor of $\sim$100--1000 compared to the typical interstellar medium. No model currently explains this suppression. We show that cosmic-ray self-confinement can significantly inhibit diffusion near pulsars. The steep cosmic-ray gradient generates Alfv{\'e}n waves that resonantly scatter the same cosmic-ray population, suppressing diffusion within $\sim$20~pc of young pulsars ($\lesssim$100~kyr). In this model, TeV halos evolve through two phases, a growth phase where Alfv{\'e}n waves are resonantly generated and cosmic-ray diffusion becomes increasingly suppressed, and a subsequent relaxation phase where the diffusion coefficient returns to the standard interstellar value. Intriguingly, cosmic rays are not strongly confined early in the TeV halo evolution, allowing a significant fraction of injected e$^\pm$ to escape. If these e$\pm$ also escape from the surrounding supernova remnant, they would provide a natural explanation for the positron excess observed by PAMELA and AMS-02.  Recently created TeV cosmic rays are confined in the TeV halo, matching observations by HAWC and H.E.S.S. While our default model relaxes too rapidly to explain the confinement of TeV cosmic rays around mature pulsars, such as Geminga, models utilizing a Kraichnan turbulence spectrum experience much slower relaxation.  Thus, observations of TeV halos around mature pulsars may provide a probe into our understanding of interstellar turbulence. 
\end{abstract}

\maketitle

\section{Introduction}
\label{sec:introduction}
Observations by the High-Altitude Water Cherenkov (HAWC) Observatory have detected a number of bright, extended TeV \mbox{$\gamma$-ray} sources coincident with both young and middle-aged pulsars~\citep{Abeysekara:2017hyn, Linden:2017vvb, 2017ATel10941....1R}. This population complements nearly two dozen extended TeV sources detected by the High-Energy Spectroscopic System (H.E.S.S.) surrounding more-distant young pulsars~\citep{Abdalla:2017vci}. This emission is thought to be powered by the inverse-Compton scattering of ambient radiation by $\sim$10~TeV  $e^{\pm}$ pairs produced in either the pulsar magnetosphere or the surrounding pulsar-wind nebula. Due to their extended nature, these sources have been named ``TeV halos"~\citep{Linden:2017vvb}. Utilizing distances provided by the Australian Telescope National Facility (ATNF) catalog~\citep{2005AJ....129.1993M}, the high luminosities of TeV halos indicate that both young and middle-aged pulsars convert a sizable fraction, $\mathcal{O}$(10\%), of their spin-down power into  $e^{\pm}$ pairs, with a spectrum that peaks in the TeV range. Intriguingly, this indicates that pulsars provide sufficient power to produce the positron excess observed by the PAMELA~\citep{2010PhRvL.105l1101A} and AMS-02~\citep{Aguilar:2013qda} instruments~\citep{Hooper:2008kg, Profumo:2008ms, Hooper:2017gtd, Cholis:2018izy}, as well as the diffuse TeV $\gamma$-ray emission observed from the Milky Way plane~\citep{Linden:2017blp} and the Galactic center~\citep{Hooper:2017rzt}. 

Recently, the HAWC Collaboration found that the TeV halos surrounding the Geminga and Monogem pulsars\footnote{Also referred to as PSR B0656+14} extend for $\sim$5$^\circ$ ($\sim$25~pc, given distances of 250$^{+230}_{-80}$~pc and 280$^{+30}_{-30}$~pc, respectively~\citep{2012ApJ...755...39V}), and found that the morphology of both halos is consistent with cosmic-ray diffusion from a central source~\citep{Abeysekara:2017old, Lopez-Coto:2017pbk}. These extensions are compatible with the broader TeV halo population, but are not anticipated in theoretical models. On the one hand, TeV halos are significantly larger than x-ray pulsar wind nebulae (PWN), which, e.g., in the case of Geminga, extends for $\lesssim$~0.05$^\circ$~\cite{Posselt:2016lot}. On the other hand, these sources are significantly smaller than the $\sim$500~pc extent expected if 10~TeV  $e^{\pm}$ diffused through the interstellar medium (ISM) with the standard diffusion constant of $\sim$3~$\times$10$^{28}$~cm$^2$s$^{-1}E_{\rm GeV}^{1/3}$ calculated from observations of local cosmic rays~\citep{Trotta:2010mx}. Studies by some of us, as well as the HAWC Collaboration, have noted that  $e^{\pm}$ diffusion within TeV halos must be significantly inhibited~\citep{Hooper:2017gtd, Abeysekara:2017old}. The most recent HAWC measurements indicate that the TeV halo diffusion constant is $\sim$4.5~$\times$10$^{27}$~cm$^{2}$~s$^{-1}$ at an energy of 100~TeV, nearly a factor of 300 lower than the average ISM.

Unfortunately, no known mechanism explains the inhibited diffusion within TeV halos. This is unlike PWN, where the morphology is explained by the formation of a termination shock when the energy density of the relativistic pulsar wind falls below the ISM density~\citep{Gaensler:2006ua}. This has motivated the HAWC Collaboration to suggest that the small diffusion constant is intrinsic to the local ISM, and that diffusion near Earth might be similarly inhibited~\citep{Abeysekara:2017old}. In contrast, it was shown that observations of $\sim$10~TeV electrons by H.E.S.S.~require that the diffusion constant near Earth is much higher, consistent with typical ISM values~\citep{Hooper:2017tkg}. This question of whether (and where) the TeV halo diffusion coefficient returns to the standard ISM value is critical for understanding the origin of the positron excess, and more broadly the characteristics of Milky Way cosmic-ray propagation. However, because the vast majority of the ISM is not in the $\sim$25~pc region observed by HAWC, TeV $\gamma$-ray observations do not provide a data-driven answer. Understanding the nature of this transition requires a physical model for TeV halo evolution.

In this \emph{paper} we show that self-generated turbulence inhibits cosmic-ray propagation near young pulsars. This builds upon previous work indicating that supernova remnants can produce localized patches of inhibited diffusion~\citep{1981A&A....98..195A, 2008AdSpR..42..486P, 2013ApJ...768...73M, DAngelo:2015cfw, 2016MNRAS.461.3552N, DAngelo:2018rou}. In particular, the steep cosmic-ray gradient produced by a single bright source generates Alfv\'en waves that propagate outward along the cosmic-ray gradient. Once excited, these Alfv\'en waves dominate the turbulence spectrum at the scattering scale because they are naturally resonant with the injected cosmic rays.
Self-generated turbulence by CR streaming is routinely taken into account while modeling cosmic-ray acceleration at the SNR shocks~\cite{1983A&A...118..223L} and propagation in the Galactic halo~\cite{1970MNRAS.147....1S,2013JCAP...07..001A}.
We apply these models to young and middle-aged pulsars, which produce a cosmic-ray population energetically dominated by $\sim$TeV leptons. In order to accurately model pulsars which are continuously injecting cosmic rays (but also rapidly spinning down), we produce the first model accounting for the coupled evolution of the $e^{\pm}$ cosmic-ray density and the self-generated turbulence.  

Our results indicate that pulsars can produce localized regions ($\sim$20~pc) where diffusion is inhibited by 2--3 orders of magnitude, consistent with TeV halo observations. Moreover, our model naturally reproduces the observed correlation between pulsar age and the size of the TeV halo. Intriguingly, our models also indicate that TeV halos do not effectively contain the majority of $e^{\pm}$ with energies below $\sim$1~TeV. This is a generic result for self-confinement in hard-spectrum sources, because the $e^{\pm}$ energy density peaks at the highest energies and makes high-energy particle confinement more efficient. Copious low-energy cosmic rays escape into the surrounding ISM early in the TeV halo evolution. Along with the high $e^{\pm}$ injection luminosity implied by HAWC and H.E.S.S. observations, this supports the idea that pulsars produce the positron excess.

\section{Cosmic-Ray Diffusion in Regions with Self-Generated Turbulence}
\label{sec:analytic}

To self-consistently model the self-generation of Alfv\'en waves and their impact on the diffusion of cosmic rays, we must solve the time-dependent propagation and diffusion of cosmic rays from an evolving source in a regime with both time- and distance-dependent diffusion. We model the time-dependent electron\footnote{Throughout this text, we will use the term ``electron" to refer to  $e^{\pm}$ pairs when the implication is clear.} injection from a pulsar as
\begin{equation}
Q_e(r,p,t) = Q_0(t) \left( \frac{p}{m_e c} \right)^{-(2+\alpha)} {
\rm e}^{-\frac{p}{p_c}} \frac{{\rm e}^{-r^2/2\sigma^2}}{(2 \pi\sigma^2)^{3/2}} \,
\end{equation}
where Q$_0$(t) is determined by the pulsars'$e^{\pm}$ luminosity
\begin{equation}
L_e (t) = (4 \pi)^2 \int_0^\infty dr \, r^2 \, \int_{p_{\rm min}}^\infty dp \, p^2 T(p) Q_e(r, p, t) 
\end{equation}
and $L_e(t) = L_0 \, (1 + t/\tau)^{-2}$ is the electron injection power of a typical pulsar. We normalize this result to match the best-fit luminosity from Geminga of $\sim$2$\times$10$^{34}$~erg~s$^{-1}$ at an age of $\sim$340~kyr~\citep{Hooper:2017gtd}. The spin-down timescale is calculated using Geminga parameters and assuming that the pulsar is losing power due to dipole radiation, giving $\tau = 9~{\rm kyr}$; see Eq.~2.11 in~\cite{Hooper:2017gtd}. We assume a source size of $\sigma$~=~1~pc, based on the assumption that  $e^{\pm}$ are accelerated at the pulsar termination shock. Changing this assumption for $\sigma$ ranging from 0.1 to 10 pc only negligibly affects our results. We assume a cutoff momentum p$_c$~=~100~TeV/c, with a minimum momentum of 1~GeV/c. 

As they diffuse, electrons are cooled through a combination of synchrotron radiation and inverse-Compton scattering, with an energy-loss rate of:

\begin{equation}
\left( \frac{dp}{dt} \right)_{\rm e} = -\frac{4}{3} \sigma_T (U_\gamma + U_B) \gamma_e^2  
\end{equation}

\noindent where we assume U$_{B}$~=~0.025~eV~cm$^{-3}$ (corresponding to a coherent magnetic field B$_{0}$~=~1~$\mu$G), and a multicomponent interstellar radiation field (ISRF) composed of CMB, infrared, optical, and UV components with radiation densities of $\rho_{\rm CMB}$~=~0.26~eV~cm$^{-3}$, $\rho_{\rm IR}$~=~0.3~eV~cm$^{-3}$, $\rho_{\rm opt}$~=~0.6~eV~cm$^{-3}$, and $\rho_{\rm UV}$~=~0.1~eV~cm$^{-3}$~\citep[see e.g.][]{Porter:2017vaa}\footnote{Notice that the radiation coming from the pulsar can dominate only within $\sim 1$ pc around the pulsar.}. 
For the highest energy electrons, the inverse-Compton scattering cross section is suppressed by Klein-Nishina effects, which can be approximated via the relation:
\begin{equation}
\frac{\sigma_{KN}}{\sigma_T}~\approx \frac{45m_e^2/64\pi^2T_i^2}{(45 m_e^2/64\pi^2T_i^2) + (E_e^2/m_e^2)}
\end{equation}
where $\sigma_{KN}$ and $\sigma_T$ are the Klein-Nishina and Thompson scattering cross-sections, respectively, and T$_i$ is the temperature of each ISRF component. 

While these electrons cool, they diffuse outward from the central source. The diffusion through the interstellar medium on distance smaller than the typical coherence length of the large scale magnetic field $B_0$ ($\lesssim$100~pc~\citep{Beck:2014pma}) is usually approximated as one-dimensional along the direction of the local magnetic flux tube. Beyond the coherence scale, the diffusion become more 3D-like and the electron density rapidly drops making the self-generated turbulence completely negligible.
We assume the radius of the flux tube to be 1 pc, roughly corresponding to the size of the PWN, and we describe the particle diffusion along the flux tube using a 1D cylindrically symmetric transport equation
\begin{equation}
\label{eq:transport}
\begin{split}
\frac{\partial f}{\partial t} 
+ u \frac{\partial f}{\partial z} 
- \frac{\partial}{\partial z} \left[D(p,z,t) \frac{\partial f}{\partial z} \right] 
- \frac{du}{dz} \frac{p}{3} \frac{\partial f}{\partial p} \\
+ \frac{1}{p^2}\frac{\partial}{\partial p} \left[ p^2 \left( \frac{dp}{dt} \right)_{\rm e} f \right]
= Q_e(p,z,t) \, ,
\end{split}
\end{equation}
where $z$ is the distance from the central pulsar. The velocity appearing in the advection term, $u$, is the Alfv\'en velocity, $v_A$, which is defined in terms of the local unperturbed magnetic field and ion mass density, $n_i$ as $v_A = B_0 / \sqrt{4 \pi m_i n_i}$.

The efficiency of cosmic-ray diffusion is governed by local magnetic field turbulence and the diffusion coefficient is expressed as:
\begin{equation}
\label{eq:diffusioncoefficient}
D(p,t) = \frac{4}{3 \pi} \frac{c \, r_{L}(p)}{k_{\rm res} \mathcal{W}(z, k_{\rm res})} 
\end{equation}
where the spectral power, $\mathcal{W}$, is calculated at the resonant wave number \mbox{$k_{\rm res} = 1 /r_L(p)$}.
The equation describing wave transport along $z$ can be written as follows~\cite{DAngelo:2018rou}:
\begin{equation}
\label{eq:wavetransportequation}
\frac{\partial \mathcal{W}}{\partial t} + v_A \frac{\partial \mathcal{W}}{\partial z} = (\Gamma_{\rm CR} - \Gamma_{\rm D}) \mathcal{W}(k, z, t)
\end{equation}
Here $\Gamma_{\rm CR}$ is the growth rate due to the CR streaming instability that will be calculated in the next section, while the term $\Gamma_{\rm D}$ takes into account that these waves are also damped by a combination of ion-neutral and nonlinear wave damping. In this study, we neglect the ion-neutral damping term, as the absence of observed H$\alpha$ emission from the bow shock of Geminga indicates that the neutral hydrogen fraction surrounding the pulsar is less than $1\%$ \citep{Caraveo+2003}. We note, however, that the neutral fraction could increase at larger distances from the pulsar and contribute significantly to terminating TeV halo activity far from the pulsar source. The remaining nonlinear damping resulting from wave-wave coupling is given by \citep{Brunetti-Cassano+2004}
\begin{equation}  \label{eq:Gamma_D}
\Gamma_{\rm NLD}(k) = c_k |v_A|
\begin{cases}
  k^{3/2} \, \mathcal{W}^{1/2} & \text{(Kolmogorov)}\\
  k^{2}   \,  \mathcal{W}          & \text{(Kraichnan)}
\end{cases} 
\end{equation}
in the Kolmogorov and Kraichnan phenomenology, respectively, while $c_k \simeq 0.052$.

We note that these growth and damping rates are both time and position dependent, and are determined by the local cosmic-ray gradient and the interstellar density. 

The background turbulence responsible for standard ISM diffusion is set to provide the galactic diffusion coefficient of $\sim$4$\times$10$^{28}$ for a rigidity of 3~GV and with a Kolmogorov spectrum, i.e., $D \propto p^{1/3}$, at distances far from the source~\citep{Johannesson:2016rlh}.

We notice that Eq.~\ref{eq:wavetransportequation} does not describe the evolution of the turbulence cascade but only accounts for the corresponding growth and damping rates at each wavelength.  
A more fundamental description of the wave transport is given in terms of a diffusion term in $k$-space~\cite[see, e.g.,][]{Brunetti-Cassano+2004,Evoli+2018}:
\begin{equation*}
\frac{\partial}{\partial k} \left[ D_{\rm kk} \frac{\partial \mathcal{W}}{\partial k} \right]
\end{equation*}
which can be approximated with a damping term, where $\Gamma_{\rm D} \sim \frac{D_{\rm kk}}{k^2}$, provided that the damping timescale is faster than diffusion and losses timescales. 
Furthermore, at each wavelength the contribution to the turbulence coming from the cascade of larger wavelengths can be neglected with respect to the self-generated one if the energy density (per logarithmic scale) in accelerated particles, $\propto p^4 f(p)$, increases with energies slower than the energy density of the cascade, which is $\propto k_{\rm res} W(k_{\rm res}) \propto p^{\beta-1}$,  where $\beta= 5/3 \; (3/2)$ for the Kolmogorov (Kraichnan) cascade. This implies that the cascade contribution can be neglected only if the particle spectrum $f(p) \propto p^{-\alpha}$ has a slope $\alpha < 10/3 \; (7/2)$ for the Kolmogorov (Kraichnan) case. Because in the following we will use $\alpha$ between 3.2 and 3.5, neglecting the cascade is not completely justified.
In this respect our final results for the level of turbulence presented in Sec.~\ref{sec:results} should be regarded as a lower limit, which translates into an upper limit for the diffusion coefficient. 
Our present approach uses Eq.~\ref{eq:wavetransportequation}, rather than its more complete form including $k$-space diffusion due to computational issues, and an inclusion of the full cascade treatment will be the subject of future work.

\subsection{Self-generated turbulence produced by $e^{+}-e^{-}$ streaming}
\label{sec:amp_rate}

The growth rate of magnetic turbulence, $\Gamma_{\rm CR}$, valid in the environment surrounding an energetic pulsar is due to the self-generated Alfv\'en waves formed by the streaming of $e^{\pm}$ pairs escaping from the PWN.
In this subsection, we perform the first calculation of the self-generated turbulence near a pulsar, noting that the results are similar to those calculated by ~\cite{Achterberg1983} and~\cite{Zweibel2003} in the case of supernova remnants dominated by cosmic-ray protons. In this derivation, we closely follow the kinetic approach employed by  \cite{Amato-Blasi2009}, modifying the results to consider a scenario where only electron/positron pairs are accelerated by the PWN and move radially from it.

{Before we introduce the formalism to compute the streaming instability induced by pairs, we briefly justify why the beam, even if its total current is zero, can amplify waves in the magnetized plasma.
The main reason is that the resonant branch of the streaming instability does not depend on the current density but on the particle number density. In contrast, the nonresonant branch of the instability is related to the total current and therefore is not expected to give any contribution, as we will show in more detail below. 
 The same argument has been used, e.g., in Ref.~\cite{Hardee-Rose:1978}, to show that a relativistic $e^{\pm}$ beam produced in the pulsar magnetosphere may amplify Alfv\'en modes if the beam energy density is larger than the magnetic energy density, even if no net currents are carried by the beam in the plasma rest frame.}

In the context of kinetic plasma theory, the growth rate can be calculated through the dispersion relation of the plasma:
\begin{equation}
  \frac{c^2 k^2}{\omega^2} = 1 + \chi   \,,
\label{eq:disp_gen}
\end{equation}
where the response function, $\chi$, can be expressed in a general form as~\citep[see, e.g.][]{1973AmJPh..41.1380K}:
\begin{eqnarray}
\chi  =   \sum_\alpha \frac{4\pi^2 q_\alpha^2}{\omega} \int_0^{\infty} dp \int_{-1}^{+1} d\mu 
\frac{p^2 v(p) (1-\mu^2)}{\omega+k v(p) \mu \pm \Omega_\alpha}   \nonumber \\
\times \left[
\frac{\partial f_\alpha}{\partial p}+\left( \frac{k
	v}{\omega}+\mu\right)\frac{1}{p} \frac{\partial f_\alpha}{\partial \mu}
\right] \,. \hspace{1cm}
\label{eq:disp}
\end{eqnarray}

Here the index $\alpha$ runs over the particle species in the plasma, $\omega$ is the wave frequency corresponding to the wave number $k$, and $\Omega_\alpha=\Omega_\alpha^*/\gamma$ is the relativistic gyrofrequency of the particles of type $\alpha$, while $\Omega_\alpha^*= e B/(m_\alpha c)$ is the particle cyclotron frequency. The ambient plasma consists of two components, a background plasma containing cold protons, ions and electrons, and an injected plasma containing the relativistic $e^{\pm}$ pairs produced by the pulsar. The velocity of each species is given by $v$. Because the background plasma is nonrelativistic, $\Omega_\alpha\approx\Omega_\alpha^*$, while the injected pairs follow the relation $\Omega_{e^-}=-\Omega_{e^+}$ due to their opposite angular velocities. 

In the diffusive system that we are considering, the drift velocity can be expressed in terms of the diffusion coefficient and distribution function as
\begin{equation}
v_d = \frac{D}{f} \, \frac{\partial f} {\partial z} \,.
\label{eq:vd}
\end{equation}

We consider the rest frame where the plasma of injected pairs is isotropic while the background plasma drifts with velocity $-v_d$. We set $n_i$ and $n_e$ to be the number density of ions (protons) and electrons in the background plasma with $n_e=n_i$. We assume that the density of injected electrons and positrons is identical, and thus $n_{e^+} = n_{e^-} \equiv N_{CR}/2$. We note that this neglects the small asymmetry resulting from the Goldreich-Julian current in the pulsar magnetosphere. The four components of the particle distribution function can then be described using the following two equations:

\begin{eqnarray}
f_{e,i}(p,\mu)&=&\frac{n_{e,i}}{2 \pi p^2}\delta (p-m_{e,i} v_d) \delta (\mu-1) \\
f_{{\rm CR},e^\pm}(p)&=&\frac{N_{\rm CR}/2}{4 \pi} g(p) \,.
\end{eqnarray}
The first equation represents the particle distribution functions of the nonelativistic background plasma, composed of electrons and ions, respectively. Here, $\mu$ is the cosine of the angle between the particle velocity and the drift velocity of the background plasma, $v_d$. The second equation represents the distribution function of relativistic  $e^{\pm}$. The function is normalized such that  $\int_{p_0}^{p_{\max}} dp~p^2 g(p)=1$ and $p_0$ and $p_{\max}$ are the minimum and maximum momenta of accelerated particles, respectively.

The contribution of the background plasma to the right hand side of Eq.~\ref{eq:disp} is easily found to be:

\begin{equation}
  \chi_{\rm pl} = -\frac{4\pi e^2 n_i}{\omega^2 m_i}\frac{\omega+k v_d}{\omega+k v_d\pm
	\Omega_i^*} - \frac{4\pi e^2 n_e}{\omega^2 m_e}\frac{\omega+k v_d}{\omega+k
	v_d\pm \Omega_e^*}.
\end{equation}

\noindent Calculating the contribution of injected pairs to the plasma growth rate is, instead, more complex. In its most general form, it can be written as:

\begin{eqnarray}
\chi_{CR} = \sum_{\alpha=e^\pm} \frac{\pi e^2 N_{CR}}{\omega} 
\int_{0}^{\infty} dp v(p) p^2 \frac{d g}{d p}		\nonumber \\
\times \int_{-1}^{+1} d\mu \frac{1-\mu^2}{\omega+k v(p)\mu \pm \Omega_\alpha}\ \,. 
\label{eq:inte}
\end{eqnarray}
where the integral over the variable $\mu$ can be written as:
\begin{eqnarray}
\int_{-1}^{+1} d\mu \frac{1-\mu^2}{\omega+k v(p) \mu \pm \Omega_e} =
{\cal P} \int_{-1}^{+1} d\mu \frac{1-\mu^2}{k v(p)\mu \pm \Omega_e} \nonumber \\
- i\pi  \int_{-1}^{+1} d\mu\ (1-\mu^2)\ \delta(k v \mu \pm \Omega_e)  \,, \hspace{1cm}
\end{eqnarray}
where $\cal P$ denotes the principal part of the integral and we have assumed that $\omega \ll \Omega_e$, which is reasonable due to our focus on low-frequency modes. The first integral represents the current term and is antisymmetric under the substitution  $e^+ \rightarrow e^-$, while the second integral, representing the resonant term, is symmetric under the same substitution. Thus, in the case of PWN dominated by electron and positron pairs, only the resonant term survives. Using Plemelj's formula for this term, we calculate the $e^+e^-$ contribution to the response function to be:
\begin{eqnarray}
\chi_{\rm CR} =  - i \frac{\pi^2 e^2 N_{CR}}{\omega k}
\int_{p_{\min}(k)}^\infty dp\ \frac{dg}{dp}\ \left[ p^2-p_{\min}(k)^2 \right]
\label{eq:inte1}
\end{eqnarray}
where we have set the minimum momentum \mbox{$p_{\min}(k)=m_e |\Omega_e^*|/k$},  which stems from the condition that the second integral in Eq.~\ref{eq:inte} is non--zero only when $\vert \mu \vert \leq 1$. This sets a limit on the particle velocity:
\begin{equation}
v(p) \geq \frac{|\Omega_e^*|}{k\gamma} 
\Longrightarrow p = \gamma m_i v(p) \geq p_{\min}(k).
\end{equation}
\noindent where $p_{\min}$ can be physically interpreted as the minimum momentum of an electron that can have a resonant interaction with a wave of a given frequency.

For the low frequencies that we are interested in, \mbox{$\omega+k v_d \ll \Omega_i^*\ll \vert\Omega_e^*\vert$}. Thus, the contribution of the background plasma term can be Taylor expanded and the constant in the dispersion relation (displacement current) neglected. So the dispersion relation reads:
\begin{equation}  \label{eq:disp1}
v_A^2 k^2 = \tilde\omega^2 - i \frac{N_{CR}}{n_i}(\tilde\omega-k v_d)\Omega_i^*  \, I_2(k) \,,
\end{equation}
where $v_A=B_0/\sqrt{4\pi m_i n_i}$ is the Alfv\'en speed, $\tilde\omega=\omega+k
v_d$ is the wave frequency in the reference frame of the relativistic pairs and we
have introduced the term $I_2$, defined as
\begin{equation}
I_2(k) = \frac{\pi}{4} p_{\min}(k) \int_{p_{\min}(k)}^{\infty} dp
\frac{dg}{dp} \left[ p^2-p_{\min}(k)^2 \right].
\label{eq:Ia2}
\end{equation}

\begin{figure*}[tbp]
\includegraphics[width=.48\textwidth]{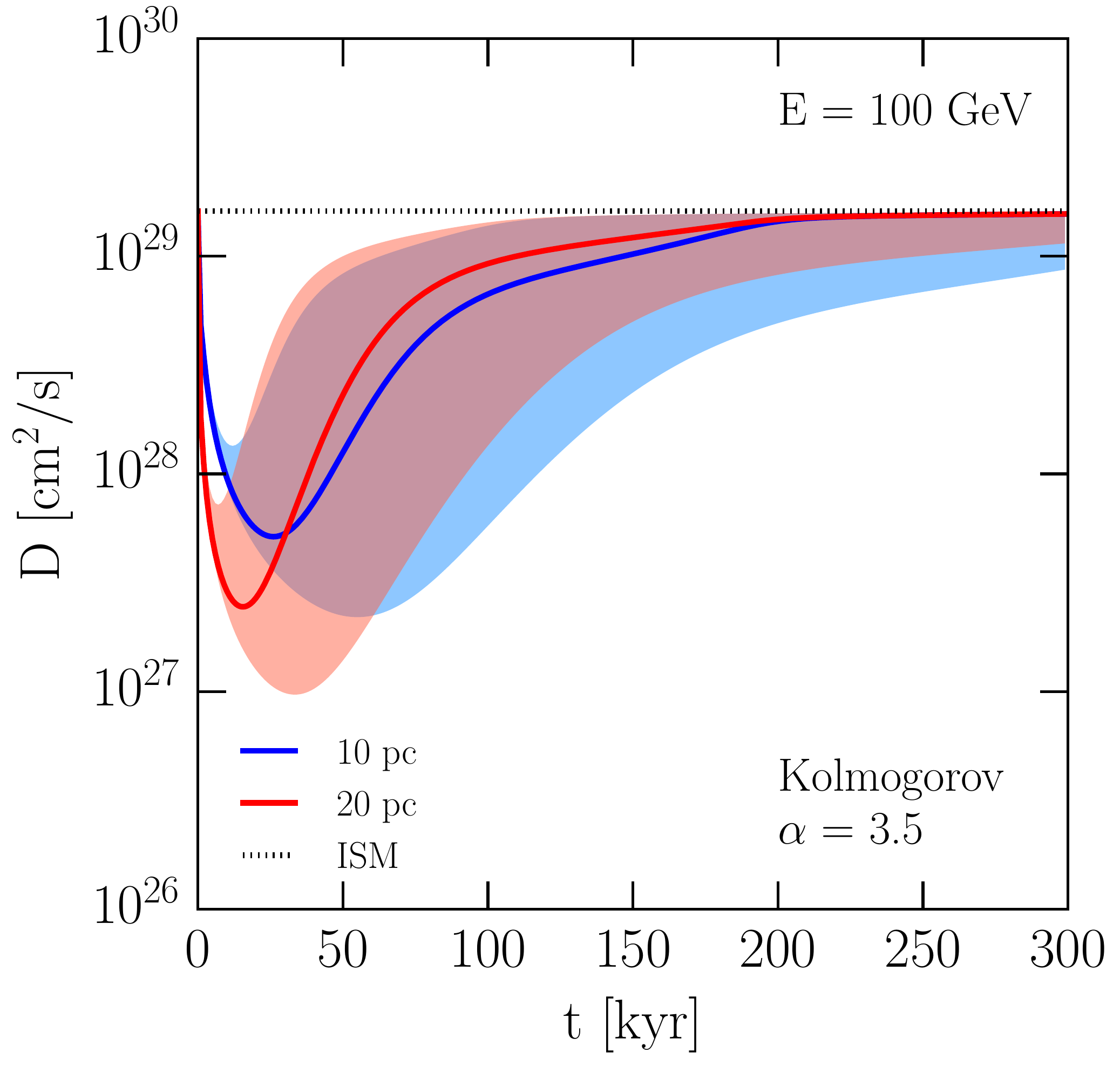}\hspace{\stretch{1}}
\includegraphics[width=.48\textwidth]{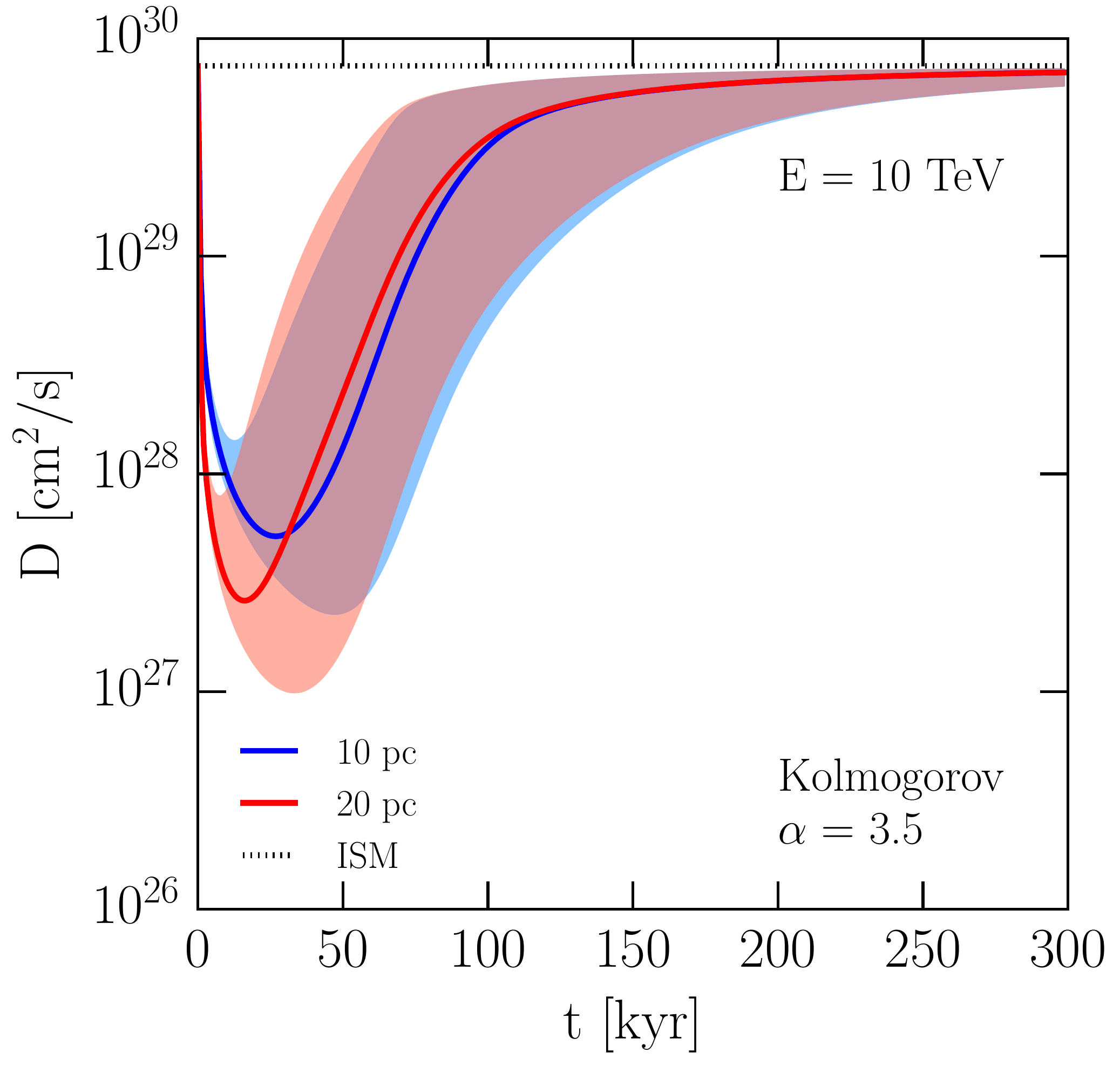}
\includegraphics[width=.48\textwidth]{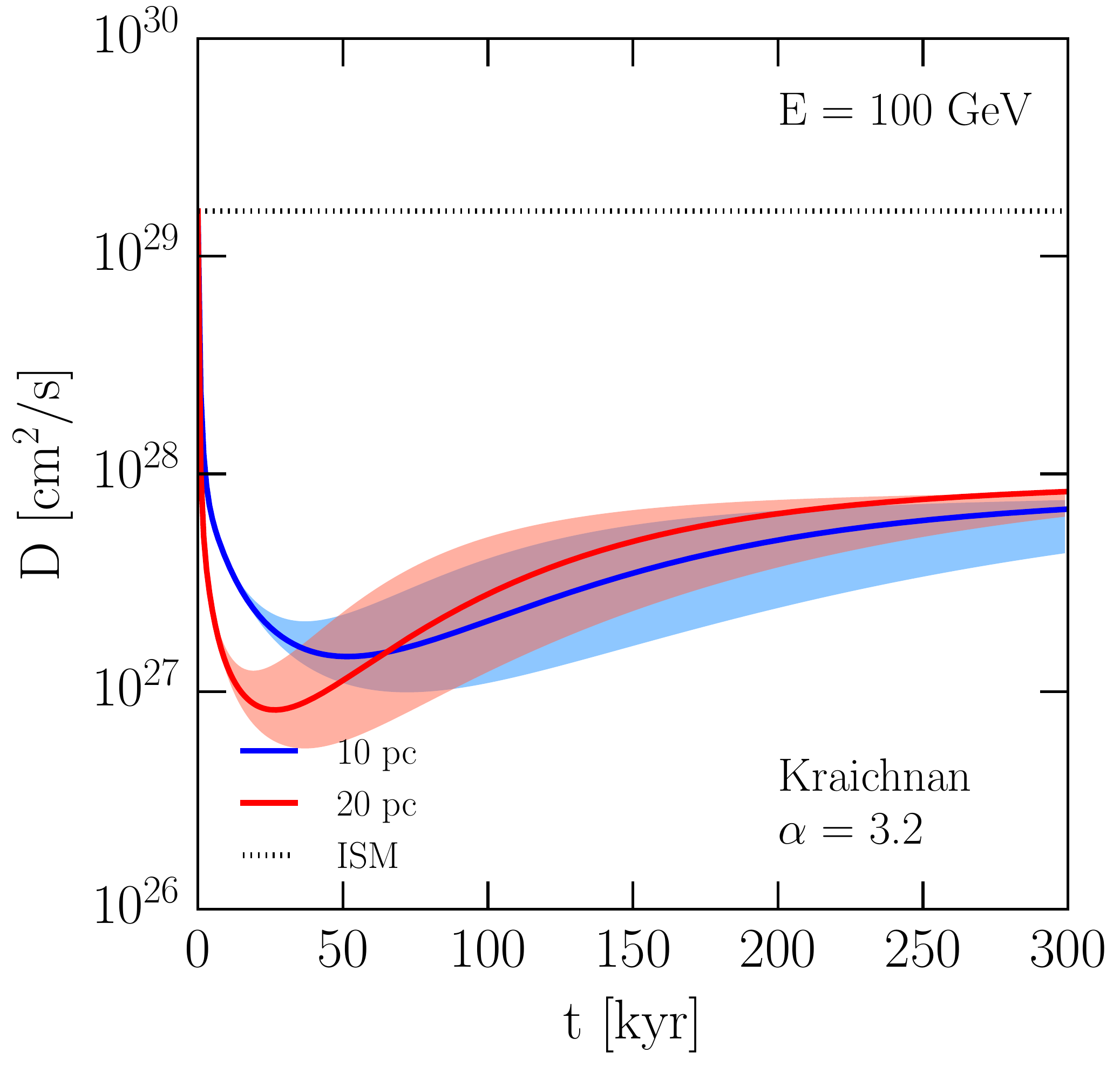}\hspace{\stretch{1}}
\includegraphics[width=.48\textwidth]{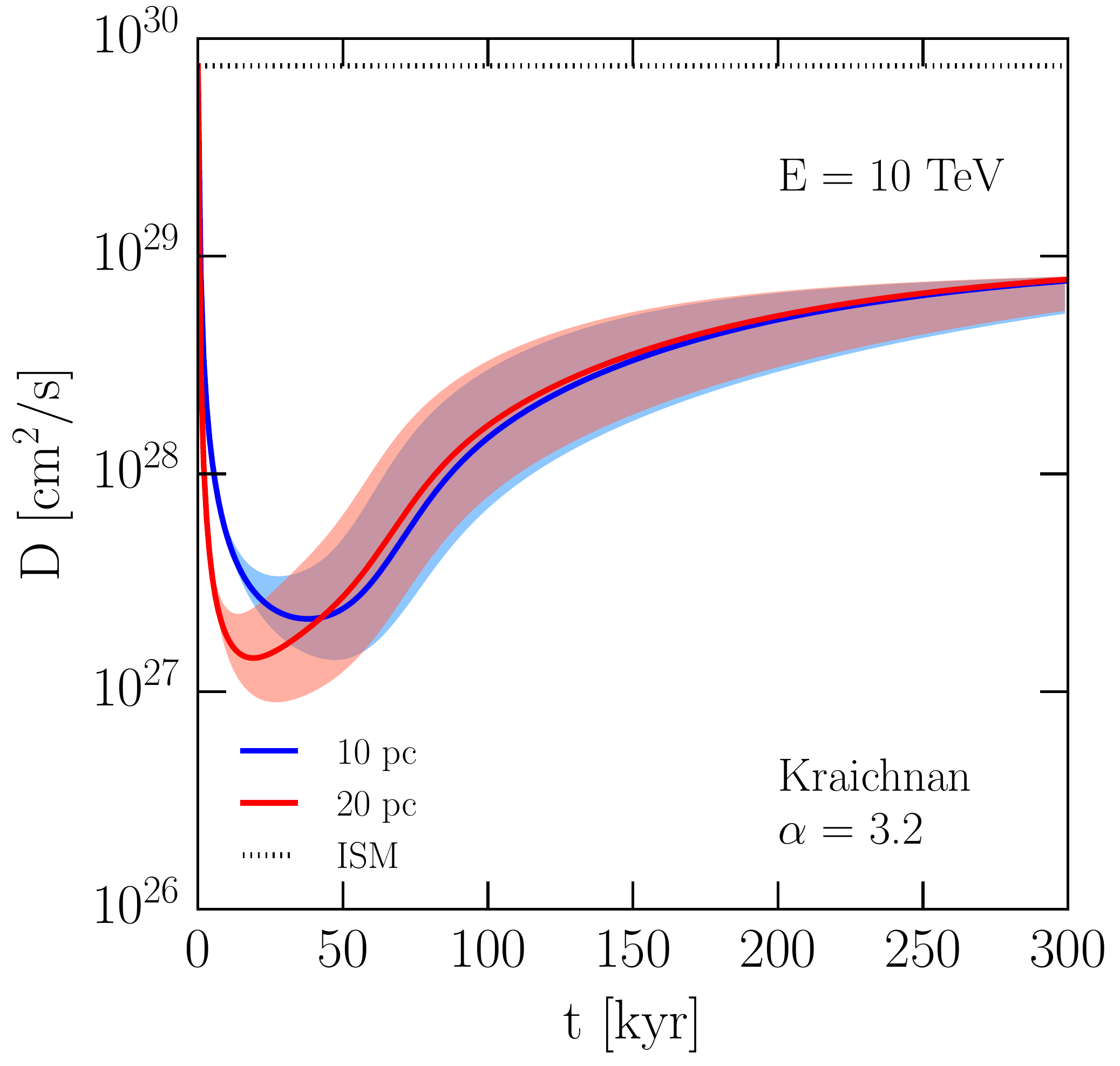}
\caption{The diffusion constant at distances of 10~pc (blue) and 20~pc (red) as a function of the pulsar age at energies of 100~GeV (left) and 10~TeV (right). Results are shown for two different models of the pulsar cosmic-ray injection and ambient diffusion parameters. Our default model (top) utilizes a Kolmogorov phenomenology for non-linear damping, and uses an electron-injection spectrum with a momentum index of -3.5. Our optimistic model (bottom) utilizes a Kraichnan phenomenology, and an electron injection index of -3.2. The solid lines utilize a background magnetic field strength of 1~$\mu$G, and the shaded region represents a range of 0.5--2~$\mu$G, with larger ambient magnetic fields producing faster relaxation to the background diffusion constant. In both cases, we find that cosmic-ray diffusion is significantly inhibited at all energies for a period between 20~kyr to 50~kyr after pulsar formation. While models utilizing Kolmogorov models and strong magnetic fields relax to standard diffusion parameters within $\sim$100~kyr, models utilizing the Kraichnan models produce inhibited diffusion through the end of our 300~kyr simulations.}
\label{fig:tevhalodiffusion}
\end{figure*}

The phase velocity of the waves in the plasma frame is $v_\phi=\tilde\omega/k$. Because we want to concentrate on waves with a velocity much smaller than the drift velocity $v_d$, we set a condition on the real part of the frequency $\tilde\omega_R \ll k v_d$. Later, we show that in the test-particle case $\tilde\omega_R = k v_A$, hence the previous condition can be expressed as $v_d \gg v_A$. We discuss the validity of this assumption at the end of the section. In this limit we can write the dispersion relation (Eq.~\ref{eq:disp_gen}) as:
\begin{equation}
v_A^2 k^2 = \tilde\omega^2 + i \, \frac{N_{CR}}{n_i} k v_d \Omega_i^* I_2(k) \,.
\label{eq:disp2}
\end{equation}
We note that Eq.~\ref{eq:disp2} is similar to the dispersion relation obtained by \cite{Amato-Blasi2009} and previously by \cite{Zweibel1979,Achterberg1983}. However in this case the current term is missing because the total current due to $e^+ e^-$ pairs is zero and there is no compensating current induced in the background plasma.

Now we look for a solution for Eq.~\ref{eq:disp2} when the non thermal spectrum can be described by a simple power law with slope $\alpha$ between $p_0$ and $p_{\max}$. The normalized distribution function $g(p)$ is then
\begin{equation}
g(p) = \frac{\alpha-3}{p_0^3} \left( \frac{p}{p_0}\right)^{-\alpha}\ \Theta(p-p_0)\ 
\Theta(p_{\max}-p)
\label{eq:gp}
\end{equation}
where $\Theta$ is the step function. The integral $I_2$ in  Eq.~\ref{eq:Ia2} can be integrated by parts, using the substitution \mbox{$s= p/p_{\min}(k) = (p/p_0) \, \ko$},  where $r_{L,0}$ is the Larmor radius of particles with momentum $p_0$, obtaining
\begin{equation}
I_2 (k) = \frac{\pi}{4} \frac{p_0^3}{(\ko)^3}
\left\{
\left[ g(s) (s^2-1) \right]_1^\infty - 2\int_1^\infty ds\ s\ g(s)
\right\}
\label{eq:ints}
\end{equation}

Note that the term in the square brackets evaluated between 1 and $\infty$ gives zero. Using the expression for $g(s)$ from Eq.~\ref{eq:gp} one finds
\begin{equation}
I_2 (k) = -\frac{\pi}{2} \frac{\alpha-3}{\alpha-2}
\left\{
\begin{array}{ccc}
(\ko)^{\alpha-3} &  & \ko \leq 1 \\
(\ko)^{-1} & & \ko \geq 1\ .
\end{array}
\right.
\end{equation}

In terms of the latter, the imaginary and real parts of the frequency can 
be written as
\begin{equation}
\tilde\omega_I^2(k) = \frac{1}{2}\left[-k^2 v_A^2 +\sqrt{k^4 v_A^4 + \sigma^2 I_2^2}\,\,  \right] 
\label{eq:omegaI}
\end{equation}
\begin{equation}
\tilde\omega_R(k)=-\frac{\sigma I_2}{2 \tilde\omega_I} \,,
\label{eq:omegaR}
\end{equation}
where $\sigma=\frac{N_{\rm CR}}{n_i}k v_d \Omega_i^*$. 

The previous equations reduce to the test-particle case when $\sigma I_2 /(k v_A)^2 \ll 1$, namely when
\begin{equation} \label{eq:condition2}
\frac{N_{\rm CR}}{n_i} \ll \frac{v_A^2}{v_d c} \frac{m_i c}{p_0} \left( \frac{p}{p_0} \right)^{\alpha-4} \,.
\end{equation}

In this limit the real part reduces to $\tilde\omega_R = k v_A$, clearly showing that those are Alfv\'en waves, while the imaginary part becomes
\begin{equation} \label{eq:omegaI_approx}
\tilde\omega_{I} \simeq \frac{\sigma I_2}{2 k v_A}
= \frac{\pi}{4} \frac{\alpha-3}{\alpha-2}\frac{v_d}{v_A} \frac{N_{\rm CR}}{n_i} \, \Omega_i^* (k r_{L,0})^{\alpha-3} \,.
\end{equation}

This is the standard result first obtained by \cite{Zweibel1979} and often used in the shock acceleration theory with the spectral index $\alpha=4$. 

In order to recover the{  final expression for the growth rate $\Gamma_{\rm CR} = 2 \tilde\omega_I$,  in Eq.~\ref{eq:omegaI_approx} we substitute the drift velocity from Eq.~\ref{eq:vd} writing down the diffusion coefficient as in Eq.~\ref{eq:transport}. Moreover we used the resonant condition for $k$, implying that $k r_{L,0}= p_0/p$, while the CR number density is rewritten as $N_{\rm CR}= 4 \pi/(\alpha-3) \, p_0^3 p^{\alpha}  f(p)$. The final result reads:
\begin{equation}
\label{eq:wavegeneration}
  \Gamma_{\rm CR}(k) = 
    \frac{2 \pi}{3} \frac{c |v_A|}{k \mathcal{W}(k) \, U_0} \left[ p^4 \frac{\partial f}{\partial z} \right]_{p_{\rm res}} \,,
\end{equation}
where $U_0= B_0^2/8\pi$.

Inserting this growth rate into the wave-transport equation (Eq.~\ref{eq:wavetransportequation}) and using Eq.~\ref{eq:diffusioncoefficient} allow  us to predict the diffusion coefficient due to the the self-generated turbulence.

The fulfillment of the test-particle approximation deserves a final comment. In deriving Eq.~\ref{eq:omegaI_approx} we required two conditions, $v_d \gg v_A$ and Eq.~\ref{eq:condition2}, to be satisfied. Using the definition of drift velocity provided in Eq.~\ref{eq:vd}, we checked a posteriori that both conditions are satisfied in all numerical solutions presented in Section~\ref{sec:results}. The validity of the test-particle approximation also implies that the amplified magnetic field  is always much smaller than the background one, $\delta B \ll B_0$, or, in other words, that the diffusion coefficient is always larger than the Bohm one calculated in the background magnetic field.

\subsection{Computational modelling}
To solve the time-dependent generation, diffusion, and energy losses of this cosmic-ray population, coupled with the wave transport equation, we sample the $e^{\pm}$ density on an equidistant spatial grid with step size of $\Delta$z~=~0.1~pc, spanning -0.5$<$z$<$0.5~kpc, and we assume that the cosmic-ray density is 0 at the outer boundaries. 
The momentum grid is log-spaced from 10~GeV to 1~PeV with 32 bins/decade. 
We assume the initial condition $f(z,p,t=0)$~=~0.0 and $\mathcal{W}(z,k,t=0) = \mathcal{W}^{\rm gal}(k)$ and evolve the distribution function using a semi-implicit Crank-Nicolson scheme with a time resolution of $\Delta$~t~=~1~yr. 
After each time step, we compute the local diffusion coefficient following Eq.~\ref{eq:diffusioncoefficient} and use this to evolve $f$ in the following step. 

\begin{figure}[tbp]
\includegraphics[width=.48\textwidth]{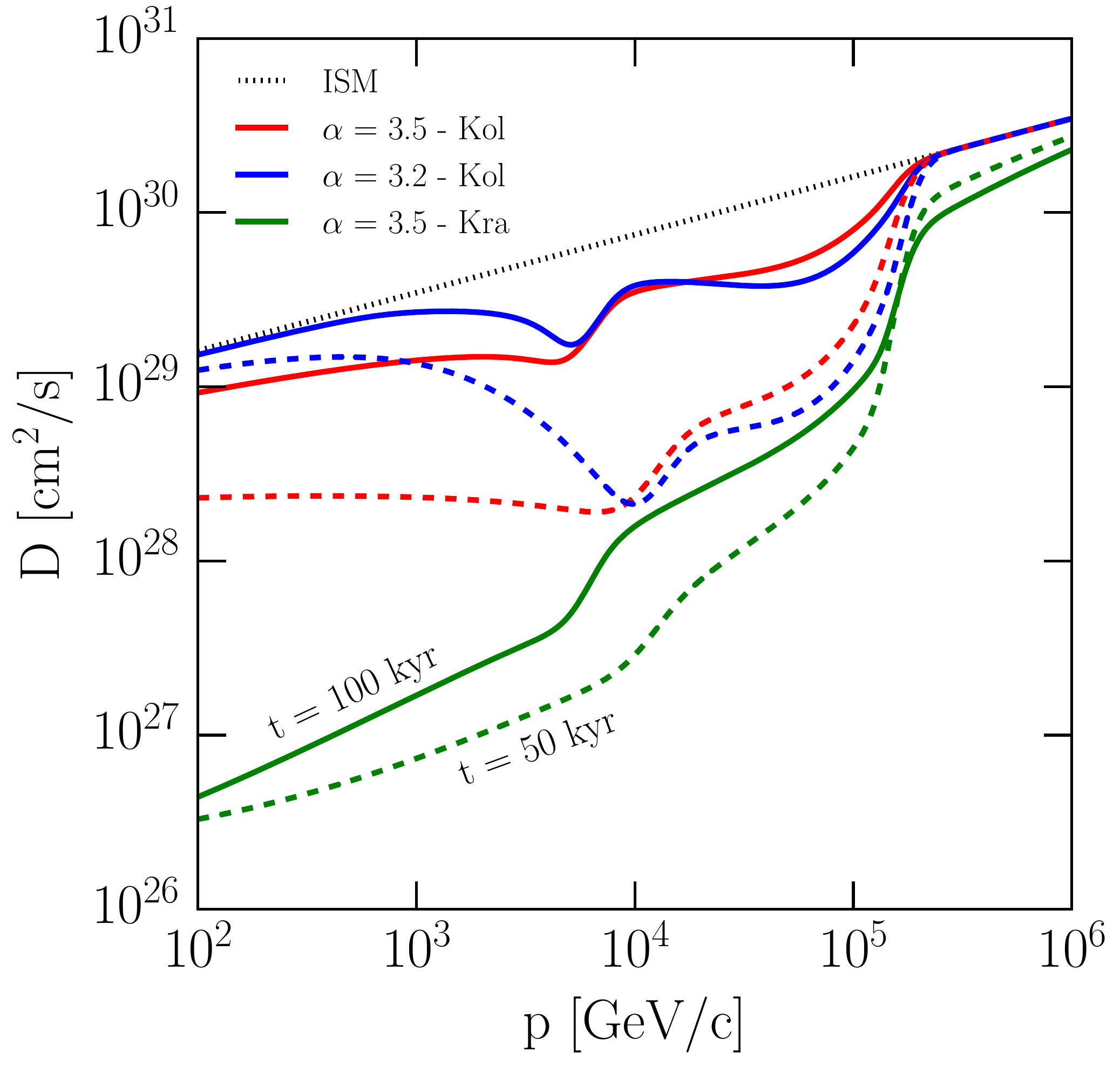}
\caption{The diffusion coefficient at a distance 10~pc from a pulsar of age 50~kyr (dashed lines) and 100~kyr (solid lines) as a function of the  $e^{\pm}$ energy. Results are shown for three models of cosmic-ray diffusion, including our default model (red lines), our optimistic model (green lines) and an intermediate model employing Kolmogorov diffusion but using a hardened  $e^{\pm}$ injection. Our results show a significant inhibition of cosmic-ray diffusion in all cases. In models with Kraichnan diffusion, the relaxation time is significantly slower, while the effective energy dependence of the diffusion constant is significantly larger.}  
\label{fig:diffusionvsenergy}
\end{figure}

\section{Results} \label{sec:results}

In Fig.~\ref{fig:tevhalodiffusion} we plot the resulting diffusion coefficient as a function of time for two different models of cosmic-ray injection and propagation. In our ``default" model, we inject cosmic-ray  $e^{\pm}$ with an injection spectrum that falls as p$^{-3.5}$, and utilize the Kolmogorov phenomenology in Eq.~\ref{eq:Gamma_D} to describe the wave damping. In our ``optimistic" model, we inject  $e^{\pm}$ with an injection spectrum that falls as p$^{-3.2}$, and utilize the Kraichnan phenomenology for wave damping. The background turbulence is always assumed to be Kolmogorov to reflect the average diffusion coefficient inferred from B/C, but our results are insensitive to this choice. In both cases, we find that the efficiency of cosmic-ray diffusion is significantly suppressed, with the most extreme effects occurring between 20 and 50~kyr after pulsar formation. At an energy of 100~GeV, the diffusion coefficient is suppressed by as much as 1.5--2 orders of magnitude in our default and optimistic models, respectively. At 10~TeV, the diffusion coefficient is suppressed even more drastically, by $\sim$2.5--3 orders of magnitude. Importantly, we find that the effective diffusion coefficient remains relatively constant over a distance exceeding 20~pc, consistent with observations of the Geminga TeV halo by the HAWC Collaboration~\citep{Abeysekara:2017old}.

The temporal evolution of our TeV halo model occurs in two phases. The first, controlled by the growth rate of  Alfv{\'e}n waves, induces a period of increasingly inhibited diffusion that propagates outward from the central source. The time-period over which the diffusion constant drops primarily depends on the characteristic spin-down time of the pulsar. During this early period, the significant injection of $e^{\pm}$ pairs results in continuous growth of the Alfv{\'e}nic turbulence. Notably, this implies that the leptons accelerated very early in the pulsars lifetime should escape by outpacing the growth of Alfv{\'e}nic turbulence. While the growth rate of Alfv{\'e}n waves, is uncertain, we note that $\sim$25\% of the total pulsar $e^{\pm}$ injection in our model occurs within the first 3~kyr after pulsar formation. The leading edge of these cosmic-ray leptons are likely to escape the TeV halo, though their fate is uncertain because such young pulsars are usually still located inside their parent SNRs and electrons must cross the remnant's envelope (presumably losing energy adiabatically) before escaping into the ISM.

\begin{figure*}[tbp]
\includegraphics[width=.49\textwidth]{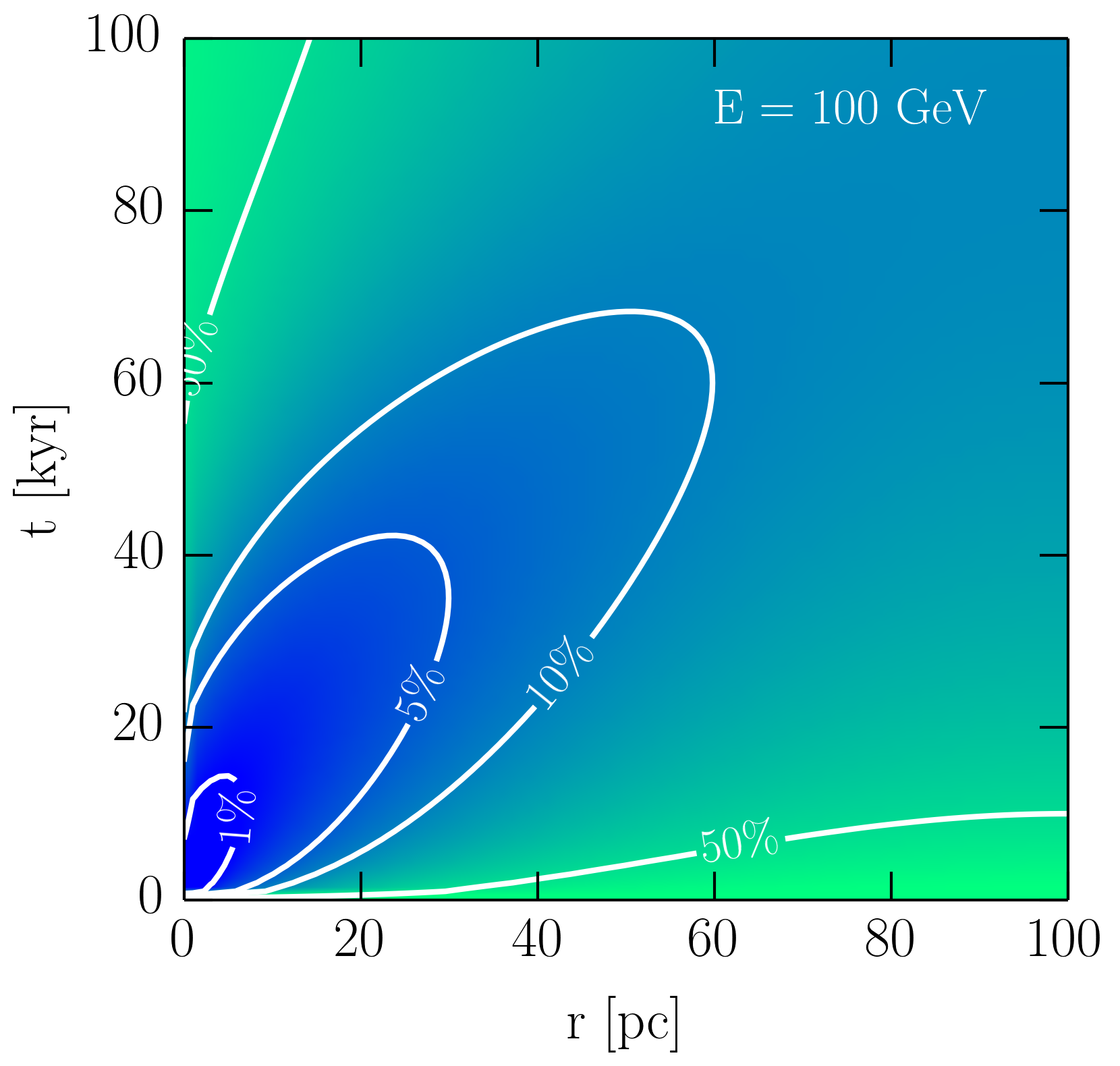}\hspace{\stretch{1}}
\includegraphics[width=.49\textwidth]{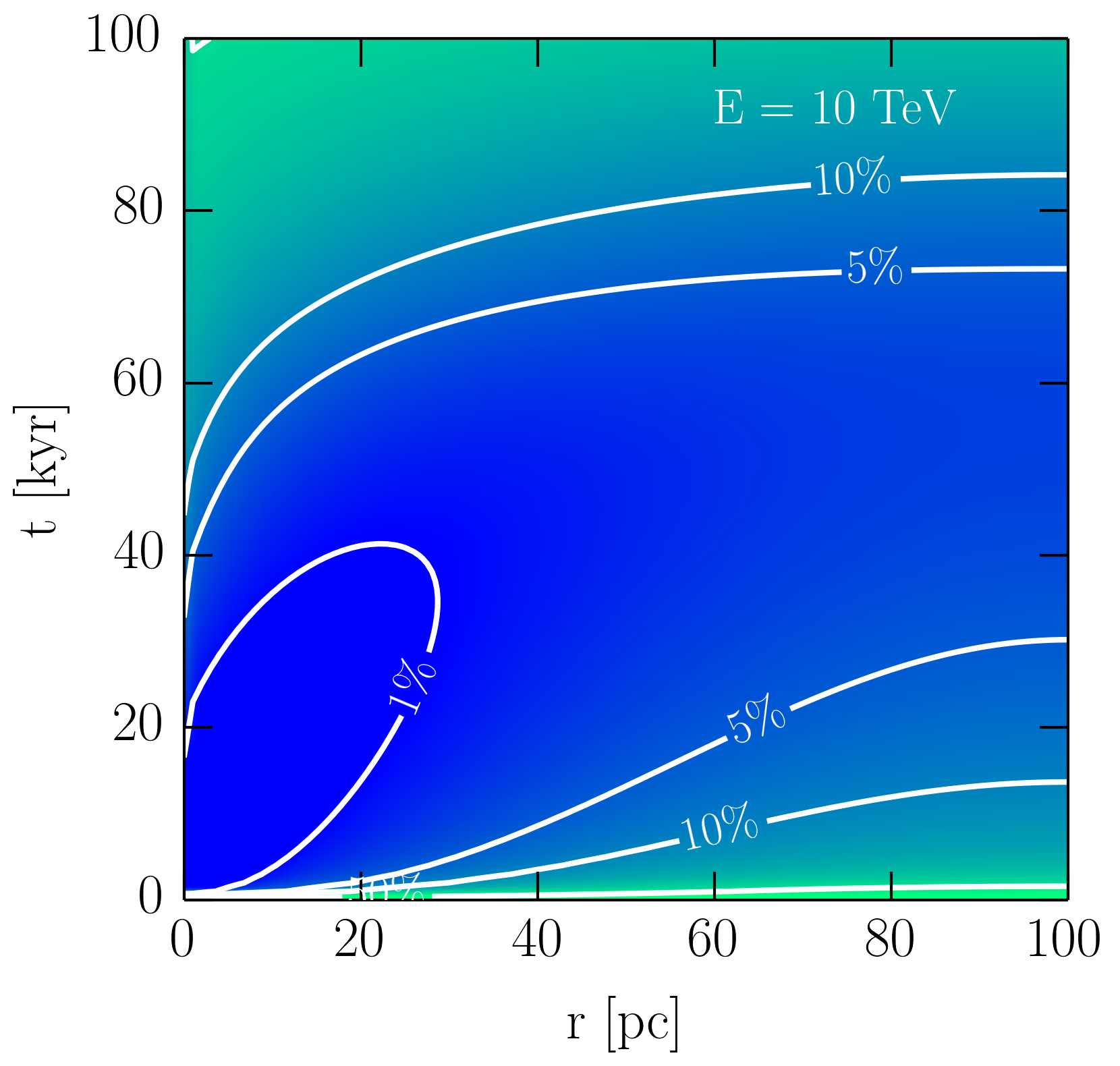}
\includegraphics[width=.49\textwidth]{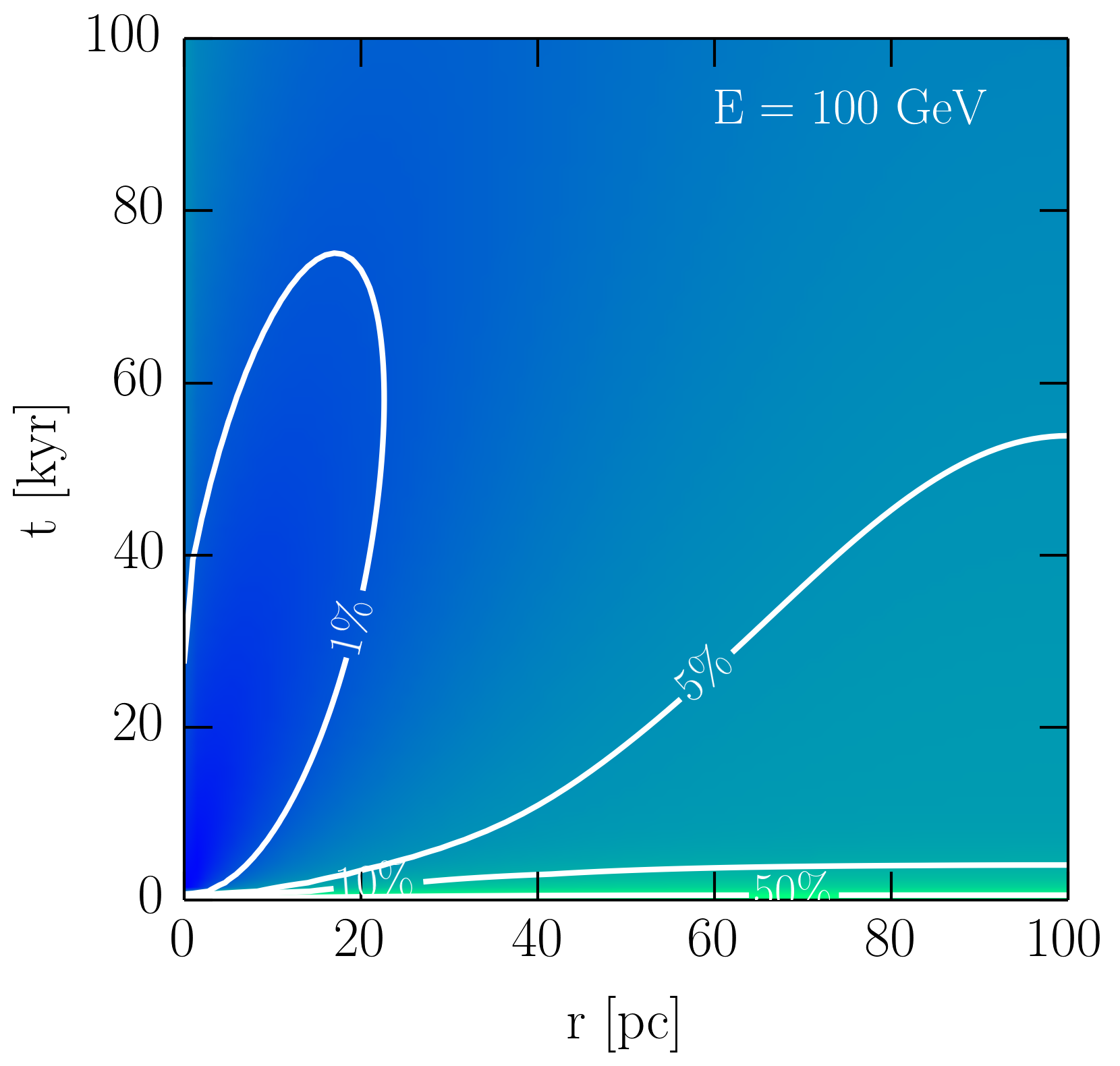}\hspace{\stretch{1}}
\includegraphics[width=.49\textwidth]{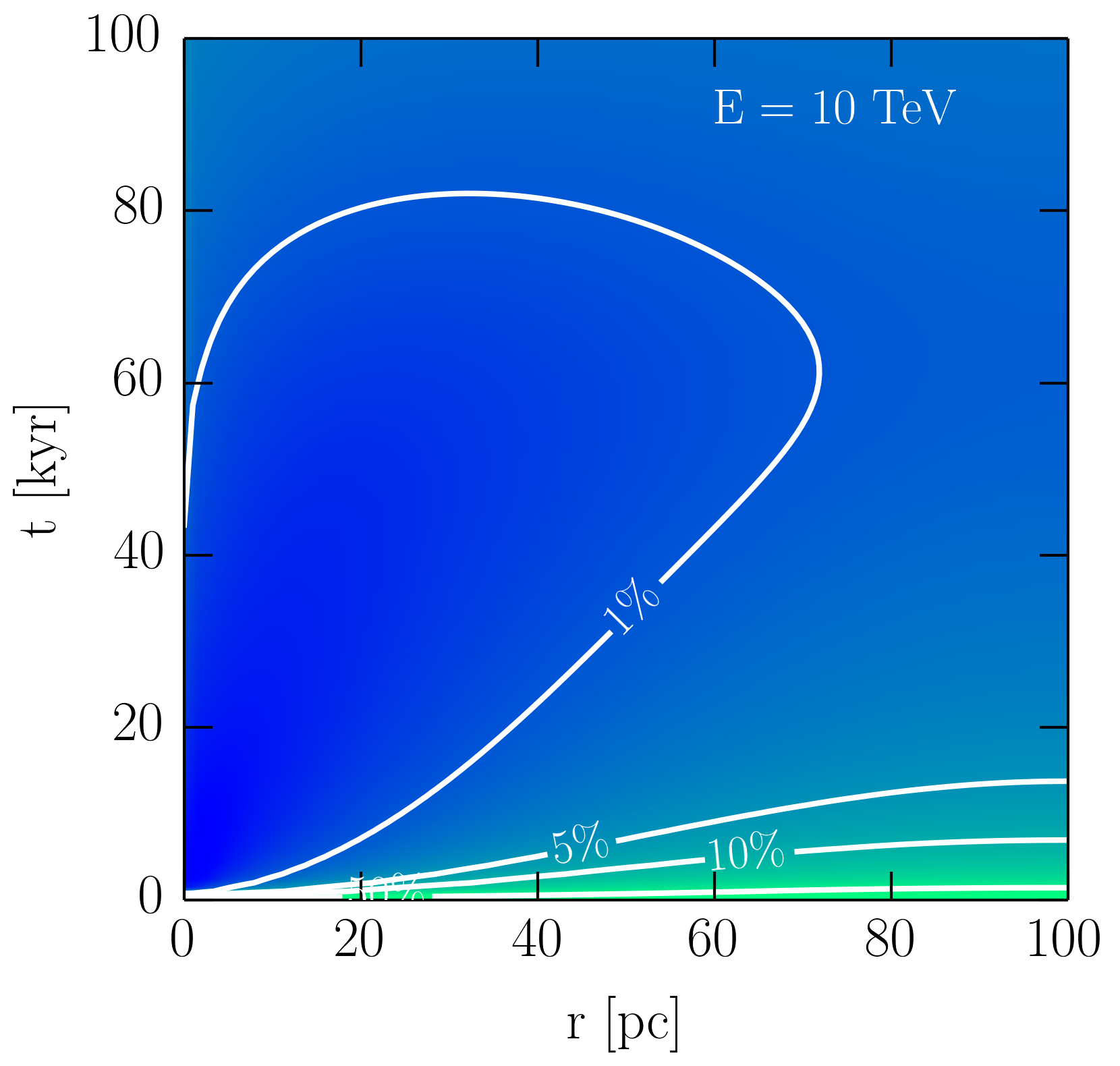}
\caption{The effective suppression of the diffusion constant (compared to the ambient background) as a function of distance (x-axis) and time (y-axis) at energies of 100~GeV (left) and 10~TeV (right) for our default (top) and optimistic (bottom) diffusion scenarios. Results are shown for a central background magnetic field strength of 1~$\mu$G. We find that significant cosmic-ray confinement persists on scales of 20--40~pc, and over time periods of 40--80~kyr. The complex spatial and temporal dependence of the inhibited diffusion, which does not monotonically return to the background value at increasing pulsar distance, explains the significant increase in the size of TeV halos over time, as observed by H.E.S.S.~\citep{Abdalla:2017vci}.}
\label{fig:tevhaloheatmap}
\end{figure*}

The second phase of TeV halo evolution begins when the pulsar luminosity decreases significantly, and the evolution is instead governed by the damping of the  Alfv{\'e}n waves and a relaxation back to standard ISM diffusion. While both our default and optimistic models behave similarly during the previous stage of turbulence growth, the relaxation rate is highly model dependent. In models with Kolmogorov damping, the TeV halo relaxes quickly back to the standard diffusion constant, typically on a timescale of $\lesssim$100~kyr. However, in models with Kraichnan damping, operating on a timescale $\tau_{\rm damp} = 1/\Gamma_{\rm damp}$, diffusion remains inhibited for much longer. This can be inferred by Eq.~\ref{eq:Gamma_D} which gives $ \tau_{\rm kol}/ \tau_{\rm kra} = (k W)^{1/2} = \delta B/B_0 \ll 1$ \cite[see also][\S~4.1]{Brunetti-Cassano+2004}. 
We name this Kraichnan model ``optimistic", because these substantially longer relaxation times provide better fits to TeV halo observations in ``mature" pulsars, such as Geminga and Monogem. Thus, the observation of significant TeV halo activity in mature pulsars may provide a powerful probe capable of constraining the physics of ISM diffusion. 

In Fig.~\ref{fig:diffusionvsenergy}, we show the energy-dependence of cosmic-ray diffusion at a distance 10~pc from a pulsar of ages 50~kyr and 100~kyr. We show results for both our default and optimistic models, as well as an intermediate result that utilizes the harder p$^{-3.2}$ cosmic-ray injection spectrum along with the default Kolmogorov turbulence model. We again note that in all cases, cosmic-ray diffusion is significantly inhibited compared to standard ISM values across a wide energy range. However, the amplitude and energy-dependence of this suppression depends sensitively on the assumed model. By showing results from an intermediate model, we separate the effects of Kraichnan diffusion and the injected electron spectrum. We find that the assumed turbulence model for damping dominates the uncertainty in our results. In addition to inducing a significantly slower relaxation time, the choice of Kraichnan diffusion significantly affects the energy-dependence of the diffusion constant. While adopting a harder cosmic-ray injection spectrum also affects the energy-dependence of diffusion, H.E.S.S. and HAWC observations of TeV halos have already placed strong constraints on the cosmic-ray injection spectrum, and thus reasonable parameter space changes (from p$^{-3.5}$ to p$^{-3.2}$) only marginally affect our main science results. In both cases, we note that observations of the energy dependent morphology of TeV halos could constrain these input parameters.

In Fig.~\ref{fig:tevhaloheatmap}, we show the effective suppression of cosmic-ray diffusion as a function of the distance from the pulsar center and the pulsar age for both our default and our optimistic models. Intriguingly, we find that the diffusion constant does not monotonically increase at distances farther from the pulsar center. Instead, the location of the minimum diffusion constant moves outward from the pulsar center as the pulsar ages. This behavior naturally explains the positive correlation between the pulsar age and the size of the TeV halo, first identified in H.E.S.S. observations~\citep{Abdalla:2017vci}. We note that for young pulsars, the TeV halo can affect the surrounding interstellar medium over regions spanning more than 100~pc. As first noted in~\citep{Hooper:2017gtd}, if TeV halos affect diffusion on 100~pc regions, and persist for 100~kyr, then approximately 2\% of the Milky Way thin disk will be filled with regions of inhibited diffusion. Because cosmic rays can remain confined in TeV halos for significantly longer than in the remaining Galactic plane, the contribution of TeV halos to Galactic $e^{\pm}$ CR diffusion and anisotropies must be carefully studied. 

\vspace{0.2cm}
\section{Discussion and Conclusions}
Recent, extremely sensitive, observations by the HAWC telescope have found intriguing evidence suggesting that the majority of young and middle-aged pulsars produce bright, spatially extended TeV $\gamma$-ray emission~\cite{Abeysekara:2017hyn, Linden:2017vvb}. 
Inefficient diffusion has been found by~\cite{2018arXiv180704182H} also around the Vela X PWN. 
However, the morphology of this emission, which is significantly larger than the pulsar termination shock yet significantly smaller than expected from free particle diffusion in the typical ISM, had remained a mystery. 

In this paper, we have produced, to our knowledge, the first time-dependent calculation of the coupled interactions between Alfv{\'e}n generated turbulence and the diffusion of $e^{\pm}$ cosmic rays from a central source. Our results indicate that the steep cosmic-ray density produced by the pulsar source excites turbulence, which significantly inhibits the propagation of these same cosmic rays. 
The magnitude of this effect closely matches the observed diffusion coefficient calculated by the authors of both Refs.~\citep{Hooper:2017gtd}~and~\citep{Abeysekara:2017old}.
Additionally, this model naturally fits two additional observational characteristics of TeV halos, the relative energy independence of their particle diffusion~\citep{Hooper:2017gtd}, and the increasing size of TeV halos as a function of the pulsar age~\citep{Abdalla:2017vci}. 
Finally, and intriguingly, the time and spatial evolution of the diffusion coefficients produced in this model naturally confine and cool $\mathcal{O}(10~\rm TeV)$  $e^{\pm}$ within the TeV halo, while allowing $\mathcal{O}(100~\rm GeV)$  $e^{\pm}$ to escape from the TeV halo. Thus, this model is in agreement with the pulsar interpretations of the positron excess.

We note that our current analysis considers only a single pulsar residing at rest in an otherwise homogeneous ISM. Further efforts are necessary to understand the effect of pulsars in more complex environments -- including pulsars that are moving with a significant kick velocity. In particular, in the interests of computational feasibility and conceptual clarity, our models have neglected two potential effects that may either enhance or disrupt the formation of TeV halos, which we discuss below. While each effect requires further investigation, we stress that they are primarily important in either very young, or very large, systems -- and are not expected to qualitatively affect our analysis of TeV halos near middle-aged pulsars. 

First, our model has neglected contributions from the parent supernova remnant. The freshly accelerated cosmic-ray protons from the SNR could further increase the cosmic-ray gradient and amplify the turbulence, particularly at low-energies where the relatively soft-spectrum cosmic-ray protons dominate the cosmic-ray energy budget. Alternatively, the supernova remnant may enhance the ambient ISM magnetic field~\citep{Reynolds:2011nk}, cooling the electron population and decreasing the cosmic-ray gradient in the TeV halo. This contribution is likely to be most important for relatively young pulsars (age $\lesssim$ 50~kyr) that are still located within their SNR. A subset of SNR, which are presently interacting with molecular clouds and have observed OH masers, have magnetic fields as high as $\sim$1~mG~\citep{1997ApJ...489..143C}. Such a significant magnetic field would cool the electron population necessary to produce the TeV halo. Alongside the presence of significant ion-neutral damping terms in the growth rate of Alfv{\'e}n turbulence, we conclude that pulsars located in dense molecular clouds are unlikely to produce TeV halo activity --- although the electrons accelerated by the pulsar may still produce significant TeV emission via interactions with the dense ISM. We note that even in regions where the parent SNR is not energetically dominant, it may affect the initial magnetic field structure, rendering our assumption of a 1D flux tube inadequate.

Second, we note that our model considers the diffusion coefficient calculated from a single TeV halo in the absence of nearby sources. However, in Fig.~\ref{fig:tevhaloheatmap}, we show that the effects of a single TeV halo may persist up to $\sim$100~pc from the pulsar. In this case, a significant fraction of the Milky Way may be affected by TeV halos, especially in dense regions along the Galactic plane. Assuming a Milky Way supernova rate of $\sim$0.02~yr$^{-1}$~\citep{Diehl:2006cf}, a supernova spatial distribution given by~\citep{Green:2015isa}, and making the reasonable assumption that all supernovae produce TeV halos, the local production rate of TeV halos is 3$\times$10$^{-2}$~kyr$^{-1}$~kpc$^{-2}$. This is roughly consistent with the observations of Geminga and Monogem at distances of  $\sim$250~pc. Intriguingly, in a region with multiple proximate TeV halos, their cumulative effect on cosmic-ray diffusion is not additive. In fact, they would generally be expected to produce destructive interference.  In Eq.~\ref{eq:wavegeneration}, the generation of Alfv\'en waves is controlled by the gradient of the local cosmic-ray density from a given source. In the case where sources from different regions produce the cosmic-ray intensity, these gradients will often cancel, minimizing the impact of self-generation turbulence on cosmic-ray propagation. This effect may be particularly important in dense star-forming regions such as star-forming clusters and the Galactic center. 
 
We stress that neither of these caveats apply to typical TeV halo observations by HAWC. In particular, observations by both HAWC and H.E.S.S. have found TeV halo extensions in the range of 5--30~pc~\citep{Abdalla:2017vci, Abeysekara:2017hyn}. Over this relatively small region, interactions between multiple TeV halos are unlikely. Of the 17 TeV halos (or potential TeV halos) observed by HAWC~\citep{Linden:2017vvb}, five have characteristic ages exceeding 100~kyr, while only 5 systems have characteristic ages below 20~kyr, where SNR contamination is likely to be most important. In the case of Geminga, the prototypical TeV halo system, the pulsar has traveled more than $\sim$70~pc from its birth location, further minimizing its interaction with the SNR shell. On the other hand, H.E.S.S. systems are typically younger (11 of the 14 confirmed systems have ages below 20~kyr~\citep{Abdalla:2017vci}). While the H.E.S.S.~Collaboration has worked extensively to differentiate supernova remnant emission from the leptonic halo, more work is needed to understand the effects of SNRs on the formation of these TeV halos. \newline
 
\section*{Acknowledgements}

We thank Elena Amato, Pasquale Blasi, Niccol\'o Bucciantini and Dan Hooper for helpful comments. 
C.E.~acknowledges the European Commission for support under the H2020-MSCA-IF-2016 action, Grant No.~751311 GRAPES – Galactic cosmic RAy Propagation: An Extensive Study.

\bibliography{tevhalos}

\end{document}